\documentclass[11pt,a4paper]{article}
  
                        \usepackage{pstricks}
                        \usepackage{multido}
                        \usepackage{pst-coil}
                        \usepackage{fancybox}
                        \usepackage{epsfig}
                        \usepackage{pst-grad} 
                        \usepackage{pst-plot} 
                        \usepackage{amsmath}
                        \usepackage{amsthm}
                        \usepackage{amssymb}
                        \usepackage{amsfonts}
                        \usepackage{verbatim} 
                        \usepackage{multicol}
\usepackage{a4wide}

\date{}

                        \newcommand{\card}[1]{\left| #1 \right|}

                        \newcommand{\loc}{\mbox{\sc loc}}
                        \newcommand{\Nloc}{\overline{\mbox{\sc loc}}}
                        \newcommand{\id}{\mbox{\sc id}}
                        
                        \newcommand{\cov}{\mbox{\sc cov}}
                        \newcommand{\Ncov}{\overline{\mbox{\sc cov}}}  
                \newcommand{\cc}{{\cal C}}
                \newcommand{\cw}{{\cal W}}

\newcounter{bob}
\newcounter{compteurTheo}

                        \newtheorem{theorem}[bob]{Theorem}
                        \newtheorem{definition}[bob]{Definition}
                        \newtheorem{lemma}[bob]{Lemma}
                        
\begin{document}

\title{Watching Systems in graphs: an extension of Identifying Codes}
\author{
{\bf David Auger$^*$, Ir\`ene Charon$^*$,}\\{\bf Olivier Hudry}\thanks{Institut
T\'el\'ecom - T\'el\'ecom ParisTech \& Centre National de la Recherche Scientifique - LTCI
UMR 5141, 46, rue Barrault, 75634 Paris Cedex 13 - France} , {\bf Antoine
Lobstein}\thanks{Centre National de la Recherche Scientifique - LTCI UMR 5141 \&
Institut T\'el\'ecom - T\'el\'ecom ParisTech, 46, rue Barrault, 75634 Paris Cedex 13 -
France}\\
{\it \footnotesize$\{$david.auger, irene.charon, olivier.hudry,
antoine.lobstein$\}$@telecom-paristech.fr}
}

\maketitle 

\medskip

\begin{center}{\small {\bf Abstract}}\end{center}
\begin{quote}
{\small We introduce the notion of watching systems in graphs, which is a generalization of that of identifying codes. We give some basic properties of watching systems, an upper bound on the minimum size of a watching system, and results on the graphs which achieve this bound; we also study the cases of the paths and cycles, and give complexity results.}
\end{quote}

\bigskip


\pagebreak
\section{Introduction and definitions}
\subsection{Identifying systems}
Many search problems, either mathematical problems or `real life' issues, come down to determine whether a particular item lies in a given set~$X$ of possible locations, and locate it if this is the case, by asking questions about its location. Let us suggest a simple model for this, in the so-called {\it non-adaptive} case, when all questions must be prepared in advance, before getting the answers.

Consider a finite set $X$ and assume that we can only query whether the item lies in certain sets $S \subseteq X$ that belong to a given family ${\cal S}$ of subsets of~$X$. For $x \in X$, the $\cal S$-{\it identifying set}, or $\cal S$-{\it label} (or simply {\it label} if there is no ambiguity) of~$x$ is the set
$$L_{\cal S}(x) = \{ S \in {\cal S} : x \in S \}.$$
We say that $\cal S$ is an {\it identifying system} of~$X$ if the labels of the elements of~$X$ are all nonempty and pairwise distinct. 

In this case, we can simply ask if the item belongs to~$S$ for every $S \in \cal S$: either all the answers will be negative and the item cannot be in~$X$, or the set of questions with positive answers will correspond to the label $L_{\cal S} (x)$ of the location~$x$ where the item is located. Since $\card{\cal S}$ can be much larger than the minimum number of questions required to always succeed, an interesting problem is to find an identifying system $\cal S'$ with $\cal S' \subseteq S$ and with minimum size.

Let us mention graph theoretical problems that are particular instances of this general framework. Karpovsky, Chakrabarty and Levitin introduced the notion of {\it identifying codes} in~\cite{karp98a}; here, with the previous notation, $X$~is the set of vertices of a finite, (in general) undirected graph and $\cal S$~is the set of all the closed neighbourhoods of the vertices of the graph (see the next section for details). More generally, with the so-called $(r,\leq \ell)$-identifying codes, one can identify sets of vertices within a certain distance (see for instance \cite{laih01a}, \cite{laih08b} or~\cite{monc06a}). Honkala, Karpovsky and Litsyn, as well as Rosendahl, studied the identification of vertices and edges of a graph using cycles (see \cite{honk01c}, \cite{honk04b}, \cite{rose03a}, \cite{rose04a}). In~\cite{honk03}, Honkala and Lobstein considered the identification of vertices in~$Z^2$, using arbitrary subsets of~$Z^2$: we shall see below that this approach is quite close to the notion of watching system. Charbit, Charon, Cohen, Hudry and Lobstein studied the general problem of identifying systems in a bipartite graph framework (\cite{Echa06}, \cite{Echa09}, \cite{char08}, \cite{char08c}). In this paper, we will introduce a new problematics in graphs, which extends the concept of identifying codes, and which can be thought of as identifying vertices with {\it subsets} of the closed neighbourhoods of the vertices of the graph.
\subsection{Notation}
We use standard notation: by {\it graph} we mean a simple, finite, undirected, generally connected, graph (if the graph is not connected, we can consider separately its connected components). If $G$ is a graph, we denote its vertex set by~$V(G)$ and its edge set by~$E(G)$. The {\it closed neighbourhood} $N_G[v]$ of a vertex~$v$ consists of $v$ and its neighbours in~$G$. For $r\geq 0$ and $v \in V(G)$, the {\it ball of radius}~$r$ {\it and centre}~$v$ is the set $B_G(v,r)$ of all vertices $x \in V(G)$ satisfying $d_G(v,x) \leq r$, where $d_G$ is the usual distance in~$G$. Obviously, $B_G(v,1)=N_G[v]$. For standard notions such as degree, diameter, spanning tree, etc., we refer to~\cite{berg83} or~\cite{BondyMurty}, whereas for the notion of {\it NP}-completeness and general background about algorithmic complexity we refer to~\cite{bart92} or~\cite{GareyJohnson}.

\subsection{Identifying codes}
Identifying codes were introduced in \cite{karp98a} in 1998 with the original motivation of fault detection in multiprocessor systems. If $G$~is a graph, an {\it identifying code} is a subset $\cc \subseteq V(G)$ such that the family
$$\{ N_G[v] : \ v \in \cc \}$$
is an identifying system of $V(G)$. The elements of $\cal C$ are usually called {\it codewords}.

Of course, such a system will exist if and only if the family of all closed neighbourhoods $\{N_G[v] : v \in V(G)\}$ is itself identifying, which means in this case that distinct vertices must have distinct closed neighbourhoods; a graph with this property is called {\it twin-free} or {\it identifiable}.

As aforementioned, in the original motivation the graph models a finite network of processors, and codewords correspond to processors equipped with a monitor able to detect a faulty processor in the closed neighbourhood of its location. Then, if there is at most one fault in the network and if every monitor sends a one-bit message referring to whether it detects a fault or not, we will be able to tell if there is a faulty processor in the graph, and locate it. See the graph~$G_1$ on Figure~\ref{ex1IDcode} for an example; one can check that a minimum identifying code in this graph has five codewords. Another example is the graph~$G_2$, depicted on Figure~\ref{ex2IDcode}, which is a star on $15$ vertices. One can check that the minimum size of an identifying code in~$G_2$ is~$14$. 

\subsection{Watching systems}
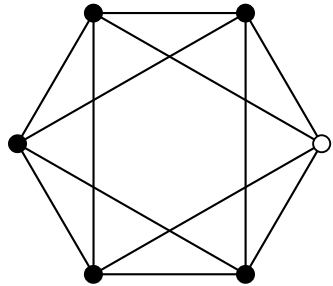
\begin{figure}[!h]
                                                \centering                                        
                                 \begin{pspicture}(-2,-2)(2,2)
                                                                     \SpecialCoor{                                    
\psline[origin={0,0}](2;0)(2;60)(2;120)(2;180)(2;240)(2;300)(2;360)%
\psline[origin={0,0}](2;0)(2;120)(2;240)(2;360)
\psline[origin={0,0}](2;60)(2;180)(2;300)(2;60)
\psdots[dotsize=0.25](2;0)(2;60)(2;120)(2;180)(2;240)(2;300)
\psdots[dotsize=0.2,linecolor=white](2;0)
} 
                                 \end{pspicture} 
                                                
 \caption{\small The graph $G_1$ and a minimum identifying code, of size five. Codewords are in black.}
                                \label{ex1IDcode} 
                                \end{figure}

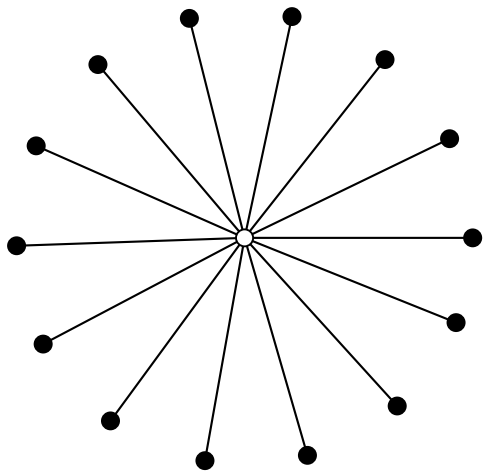
\begin{figure}[!h]
                                                \centering                                        
                                                \begin{pspicture}(-2,-4)(2,4)                                           
                                                        \SpecialCoor{
                                                \multido{\i=0+26}{14}{%
                                                \psline(0;0)(3;\i)
                                                \psdot[origin={0,0},dotsize=0.25](3;\i)%
                                                }                                               
                                                \psdot[origin={0,0},dotsize=0.25](0;0)
                                                \psdot[origin={0,0},dotsize=0.2,linecolor=white](0;0)
                                                 }                                           
                                                \end{pspicture} 
                                                
\caption{\small The graph $G_2$ and a minimum identifying code, of size $14$. Codewords are in black.}
                                \label{ex2IDcode} 
                                \end{figure}

The graph $G_1$ (Figure~\ref{ex1IDcode}) is slightly pathological, because requiring five codewords to monitor six vertices is very much to ask (in fact, $n-1$ codewords is the maximum that can be required for a graph on $n$ vertices, see \cite{char03c} or~\cite{grav04}). The reason why we need so many codewords is that the closed neighbourhoods of two distinct vertices only differ by at most two vertices (the same phenomenon is also true for the leaves in~$G_2$ on Figure~\ref{ex2IDcode}), and in this context a codeword has no choice but to check its whole closed neighbourhood, so that two distinct codewords check almost the same sets of vertices.

There are problems in which this situation is close to reality. For instance, consider a smoke detector: it has no choice but to detect smoke, regardless of the direction where it came from.
So an identifying code is a good model for a fire-monitoring system in a building.

On the other hand, for instance in fault detection in multiprocessor systems, it seems plausible that we could easily assign a smaller control area to every detector by simply {\it not connecting it} to some adjacent vertices. Let us use the term {\it watcher} instead of codeword for this generalization.

First, let us define it informally with two examples. Assume that an edge between two vertices $a$ and~$b$ denotes the possibility for a watcher in~$a$ to watch out what happens in~$b$, but that we can choose not to use this possibility: thus we can assign to a watcher located at a vertex~$v$ a {\it watching zone}, which will be any subset of~$N_G[v]$. 

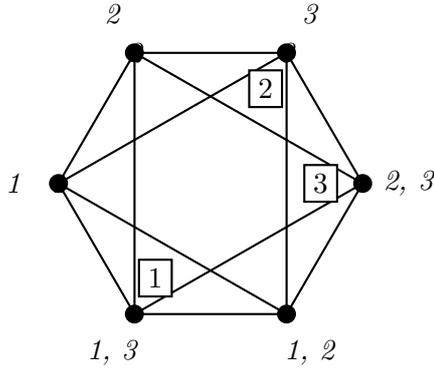
\begin{figure}[!h]
                                                \centering                                
                                                \begin{pspicture}(-2,-2)(2,2.5)                                                
                                                 \SpecialCoor{                                     
\psline[origin={0,0}](2;0)(2;60)(2;120)(2;180)(2;240)(2;300)(2;360)%
\psline[origin={0,0}](2;0)(2;120)(2;240)(2;360)
\psline[origin={0,0}](2;60)(2;180)(2;300)(2;60)
\psdots[dotsize=0.25](2;0)(2;60)(2;120)(2;180)(2;240)(2;300)
                                                \rput[origin={0,0}](2;60){\it 3}
                                                \rput[origin={0,0}](2;120){\it 2}
                                                \rput[origin={0,0}](2;180){\it 1}
                                                \rput[origin={0,0}](2.6;0){\it 2, 3}
                                \rput(1.45;0){\psframebox[fillstyle=solid,fillcolor=white]{$3$}}
                                                \rput[origin={0,0}](2.6;60){\it 3}
                                \rput(1.45;60){\psframebox[fillstyle=solid,fillcolor=white]{$2$}}
                                                \rput[origin={0,0}](2.6;120){\it 2}
                                                \rput[origin={0,0}](2.6;180){\it 1}
                                                \rput[origin={0,0}](2.6;240){\it 1, 3}  
                               \rput(1.45;240){\psframebox[fillstyle=solid,fillcolor=white]{$1$}}
                                                \rput[origin={0,0}](2.6;300){\it 1, 2}
                                                }   
                                                \end{pspicture}
                                                
\caption{\small The graph $G_1$ with a minimum watching system, of size three. Watchers' locations are written down inside squares and labels nearby vertices, in italics.}
                                \label{ex1watchers}
                                \end{figure}

Let us try this on $G_1$ and check out Figure~\ref{ex1watchers}: we only need three watchers with this protocol (the locations of the watchers $1$, $2$ and $3$ are written down in squares, whereas the label of each vertex, i.e., the set of watchers watching it, is written down in italics nearby, so that the watching zone of each watcher can be retrieved), when five codewords were needed previously. All we have to check is that the labels of all vertices are nonempty and different, and that each watcher only watches vertices in the closed neighbourhood of its location.

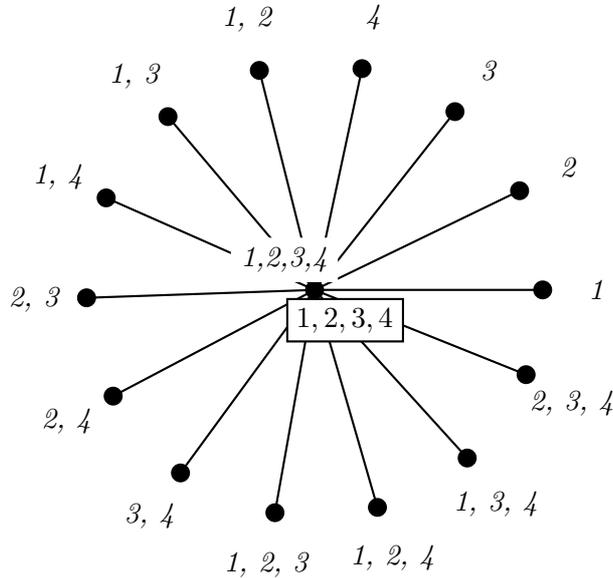
\begin{figure}[!h]
                                                \centering                                        
                                                \begin{pspicture}(-2,-4)(2,4)                                      
                                                        \SpecialCoor{                                     
                                                \multido{\i=0+26}{14}{%
                                                \psline(0;0)(3;\i)
                                                \psdot[origin={0,0},dotsize=0.25](3;\i)%
                                                }                                             
                                                \psdot[origin={0,0},dotsize=0.25](0;0)
                                                \rput(3.7;0){\it 1}        
                                                \rput(3.7;26){\it 2}
                                                \rput(3.7;52){\it 3}
                                                \rput(3.7;78){\it 4}
                                                \rput(3.7;104){\it 1, 2}
                                                \rput(3.7;130){\it 1, 3}
                                                \rput(3.7;156){\it 1, 4}
                                                \rput(3.7;182){\it 2, 3}
                                                \rput(3.7;208){\it 2, 4}
                                                \rput(3.7;234){\it 3, 4}
                                                \rput(3.7;260){\it 1, 2, 3}
                                                \rput(3.7;286){\it 1, 2, 4}
                                                \rput(3.7;0310){\it 1, 3, 4}
                                                \rput(3.7;336){\it 2, 3, 4}
                                                                }                
\rput(0.4,-.4){\psframebox[fillcolor=white,fillstyle=solid]{$1,2,3,4$}}
 \rput(-.4,.4){\psframebox[fillcolor=white,fillstyle=solid,linecolor=white]{{\it 1,2,3,4}}}
                                                \end{pspicture} 
                                                
\caption{\small A minimum watching system in $G_2$, of size four.}
                                \label{ex2watchers} 
                                \end{figure}

What we can also do is to place several watchers at the same location, with distinct watching zones. For instance consider $G_2$ (see Figure~\ref{ex2watchers}): only four watchers are needed whereas $14$ codewords were necessary. This can be thought of as a single detector in the centre of the star, but needing four bits instead of one to send information, since it has $15$ different vertices to watch. Thus watchers also enable us to model a monitoring system where monitors could simply tell where they detect a fault, but where the cost of a monitor is proportional to the number of bits needed to send this information.

Let us define this formally:
\begin{definition}
A {\it watching system} in a graph $G=(V(G),E(G))$ is a finite set 
$$\cw=\{w_1, w_2, \ldots, w_k\}$$
where each $w_i$ is a couple $w_i=(v_i,Z_i)$, where $v_i$ is a vertex and $Z_i \subseteq N_G[v_i]$, such that $\{Z_1,\ldots,Z_k\}$ is an identifying system.
\end{definition}
\noindent We will often represent watchers simply by integers, as we did in Figures~\ref{ex1watchers} and~\ref{ex2watchers}. Note that any graph $G$ admits the trivial watching system $\{(v,\{v\}): v\in V(G)\}$.

If $\cw$ is a watching system in $G$ and $w=(v,Z) \in \cw$ is a watcher, we will say that $v$ is the {\it location} of~$w$, or that $w$ is {\it located} at~$v$. The set~$Z$ is the {\it watching zone}, or {\it watching area}, of~$w$, and if $x \in Z$ we say that $w$ {\it covers}~$x$, or that $x$ is {\it covered by}~$w$. We say that $w$ {\it separates} the vertices $x$ and~$y$ (or $x$ from~$y$) if $w$ covers $x$ and does not cover~$y$, or the other way round. Therefore, $\cw$ is a watching system of~$G$ if every vertex is covered by at least one watcher in~$\cw$ and any two distinct vertices are separated by at least one watcher in~$\cw$. Let us define the $\cw$-{\it label}, or $\cw$-{\it identifying set}, or simply {\it label}, of a vertex~$v$ as the set $L_\cw(v)$ of watchers covering~$v$. We will say that a vertex~$v$ is {\it identified} by~$\cw$ if its label $L_\cw(v)$ is nonempty ($v$~is covered by one watcher at least) and there is no other vertex in~$G$ with the same label. Thus another way to express the fact that $\cw$ is a watching system is to say that all vertices in~$G$ are identified by~$\cw$.

\section{First properties of watching systems}
Let us recall that a {\it dominating set} in~$G$ is a subset $\Gamma$ of~$V(G)$ such that every vertex not in~$\Gamma$ is adjacent to at least one element in~$\Gamma$. Let respectively $w(G)$, $\gamma(G)$ and~$i(G)$ denote the minimum sizes of a watching system, of a dominating set and, when it exists, of an identifying code in~$G$. These parameters will be called {\it watching number}, {\it domination number}, and {\it identifying number}, res\-pectively.

If we have $k$ questions to be answered by yes or no, there are $2^k -1$ possibilities to answer all these questions without answering always by the negative, so we get a trivial lower bound for the size of a watching system. It is known that this bound also holds for identifying codes (see~\cite{karp98a}). Noticing that an identifying code, when it exists, defines a watching system in an obvious way, we have the following relationship involving $|V(G)|$ and the watching and identifying numbers:
\begin{theorem} \label{w(G)i(G)}
For any graph $G$, we have:
$$ \left\lceil \log_2 ( \card{V(G)} +1 )\right\rceil \leq w(G).$$
For any twin-free graph $G$, we have:
$$w(G) \leq i(G).$$
\end{theorem}
\noindent We now compare the watching and domination numbers of a graph, with the following result, where $\Delta(G)$ denotes the maximum degree of~$G$:
\begin{theorem}
For any graph $G$, we have:
$$\gamma(G) \leq w(G) \leq \gamma(G) \cdot \left\lceil \log_2 ( \Delta(G) +2)\right\rceil .$$
\end{theorem}
\noindent {\bf Proof.} If $\cw$ is a watching system, then the set of the watchers' locations in~$\cw$ is a dominating set, so we have the left-hand inequality. On the other hand, if we have a dominating set $\Gamma \subseteq V(G)$ of size $\gamma(G)$, we can identify all vertices simply by locating enough watchers at every vertex of~$\Gamma$. One just has to notice that in order to identify a vertex $v$ and its (at most) $\Delta (G)$ neighbours, we need at most $p:=\left\lceil \log_2 ( \Delta(G) +2 )\right\rceil$ watchers, since a set with $p$ elements has at least $\Delta(G)+1$ nonempty subsets. \hfill $\square$
\section{An upper bound for the watching number}
It is known that $i(G) \leq \card{V(G)} -1$ for any connected twin-free graph with at least three vertices (see \cite{char03c}, \cite{grav04}), and that this bound is reached, for instance, by the star, cf. Figure~\ref{ex2IDcode}. We prove a much better upper bound for watching systems, namely $2n/3$, in Theorem~\ref{bound}, the proof of which will use the following three lemmata.
\begin{lemma} \label{partial_graph}
Let $G$ be a graph and $H$ be a partial graph of $G$, i.e., with $V(H)=V(G)$ and $E(H)\subseteq E(G)$. Then
$$w(H) \geq w(G).$$
\end{lemma}
\noindent {\bf Proof.} If ${\cal W}$ is a watching system for~$H$, then the same~${\cal W}$ is a watching system for~$G$, since two adjacent vertices in~$H$ are also adjacent in~$G$. \qed

\medskip

\noindent Note that this monotony property does not hold in general for identifying codes.
\begin{lemma} \label{lemmefeuille}
Let $T$ be a tree, $x$ be a leaf of $T$, and $y$ be the neighbour of~$x$.

{\rm (a)} There exists a minimum watching system for~$T$ with one watcher located at~$y$.

{\rm (b)} If $y$ has degree~2, there exists a minimum watching system for~$T$ with one watcher located at~$z$, the second neighbour of~$y$.
\end{lemma}
\noindent {\bf Proof.} (a) A watching system must cover~$x$, so there is a watcher~$w_1$ located at $x$ or~$y$, with $x\in Z$. If $w_1=(x,Z)$, then we can replace it by $w_2=(y,Z)$, since $N_G[y] \supseteq N_G[x]$. 

\begin{figure}
\begin{center}
\includegraphics*[scale=0.75]{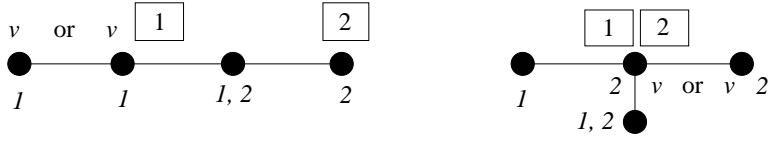}
\caption {\small Trees with four vertices.} \label{fig:order4}
\end{center}
\end{figure}

(b) If $y \notin Z$, then one other watcher must cover~$y$, and if $y \in Z$, then one must separate~$x$ and~$y$, since $x\in Z$. In both cases, the task can be done by a watcher located at~$z$. \qed
\begin{lemma} \label{lemmeOrdre4}
Let $T$ be a tree with four vertices, and let $v$ be a vertex of~$T$; there exists a set~${\cal W}$ of two watchers such that 
\begin{itemize}
\item the vertices in $V(T) \setminus \{v\}$ are covered and pairwise separated by~${\cal W}$ --- in this case, we shall say, with a slight abuse of notation, that ${\cal W}$ is a watching system of $V(T)\setminus \{v\}$;

\item the vertex $v$ is covered by at least one watcher. 
\end{itemize}
\end{lemma}
\noindent {\bf Proof.} On Figure \ref{fig:order4}, we give all possibilities: the two trees with four vertices, and for each of them, the two possible locations for~$v$ ($v$ is a leaf, or $v$ is not a leaf). \qed

\begin{figure}
\begin{center}
\includegraphics*[scale=0.8]{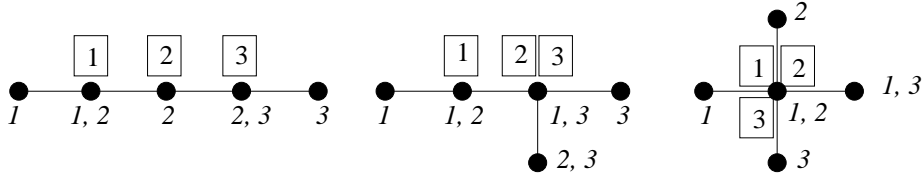}
\caption {\small The case $n$=5 in Theorem~\ref{bound}.} \label{fig:order5}
\end{center}
\end{figure}

\begin{theorem} \label{bound}
Let $G$ be a connected graph of order $n$, i.e., with $n$ vertices. 

\begin{itemize}
\item If $n=1$, $w(G)=1$.
\item If $n=2$ or $n=3$, $w(G)=2$.
\item If $n=4$ or $n=5$, $w(G)=3$.
\item If $n\notin \{1,2,4\}$, $w(G) \leq \frac{2n}{3}$.
\end{itemize}
\end{theorem}
\noindent {\bf Proof.} For $n=1$, $n=2$, or $n=3$, the result is direct. For $n=4$, it is necessary to have at least $\lceil \log_2(5) \rceil = 3$ watchers and it is easy to verify that this is sufficient. For $n = 5$, all possibilities are given by Figure~\ref{fig:order5} and we can see that we always have $w(G)=3$.

We proceed by induction on $n$. We assume that $n \geq 6$ and that the theorem is true for any connected graph of order less than~$n$. 

Let $G$ be a connected graph of order $n$. Let $T$ be a spanning tree of~$G$; we will prove that $w(T) \leq \frac{2n}{3}$ and then the theorem will result from Lemma~\ref{partial_graph}. We denote by $D$ the diameter of $T$ and we consider a path $v_0, v_1, v_2, \ldots, v_{D-1}, v_D$ of $T$, with length~$D$.

We distinguish between four cases, according to some conditions on the degrees of $v_{D-1}$ and~$v_{D-2}$.\\

\begin{figure}
\begin{center}
\includegraphics*[scale=0.8]{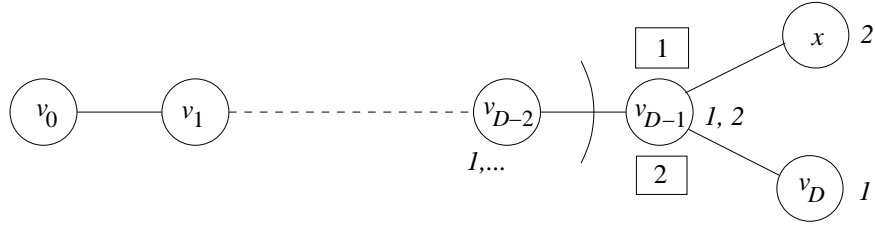}
\caption {\small First case of Theorem~\ref{bound}: the degree of $v_{D-1}$ is equal to~3.} \label{fig:firstCase}
\end{center}
\end{figure}

\noindent $\bullet$ {\it First case: the degree of $v_{D-1}$ is equal to~3}\\
The vertex $v_{D-1}$ is adjacent to a vertex $x$ other than $v_{D-2}$ and~$v_D$; because $D$ is the diameter, clearly $x$ and $v_D$ are leaves of~$T$ (see Figure~\ref{fig:firstCase}). We consider the tree obtained by removing $x$, $v_{D-1}$ and~$v_D$ from~$T$; this new tree $T'$ has order~$n-3$. 

If $n \geq 8$ or if $n=6$, we consider a minimum watching system~${\cal W}$ for~$T'$; if $n=7$, then $T'$ is of order~4, and, using Lemma~\ref{lemmeOrdre4}, we choose a set~${\cal W}$ of two watchers which is a watching system for $V(T') \setminus \{v_{D-2}\}$ {\it and} covers the vertex~$v_{D-2}$. 

Then for $T$, in both cases, we add to ${\cal W}$ two watchers $w_1 = (v_{D-1}, \{v_{D-2},$ $v_{D-1}, v_D\})$ and $w_2 = (v_{D-1}, \{v_{D-1}, x\})$. On Figure \ref{fig:firstCase}, we rename 1 and~2 these watchers. Then ${\cal W} \cup \{w_1, w_2\}$ is a watching system for~$T$. So, $w(T) \leq |{\cal W}|+2 \leq w(T')+2$.

Now we use the induction hypothesis: if $n \geq 8$ or $n=6$, then \(w(T) \leq \frac{2}{3}(n-3)+2 = \frac{2n}{3}\); and if $n=7$, then $w(T) \leq 2+2=4 < \frac {2}{3} \times 7$.\\

\begin{figure}
\begin{center}
\includegraphics*[scale=0.8]{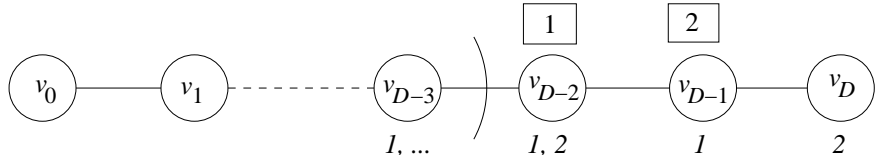}
\caption {\small Second case of Theorem~\ref{bound}: the degrees of $v_{D-1}$ and $v_{D-2}$ are equal to~2.} \label{fig:secondCase}
\end{center}
\end{figure}

\noindent $\bullet$ {\it Second case: the degrees of $v_{D-1}$ and $v_{D-2}$ are equal to~2}\\
The neighbours of $v_{D-1}$ are $v_{D-2}$ and~$v_D$, the neighbours of $v_{D-2}$ are $v_{D-3}$ and~$v_{D-1}$ (see Figure~\ref{fig:secondCase}). We consider the tree obtained by removing $v_{D-2}$, $v_{D-1}$ and~$v_D$ from~$T$; this new tree $T'$ has order~$n-3$.

If $n \geq 8$ or if $n = 6$, we consider a minimum watching system ${\cal W}$ for~$T'$; if $n = 7$, $T'$ is of order~4; again using Lemma~\ref{lemmeOrdre4}, we choose a set~${\cal W}$ of two watchers which is a watching system for $V(T') \setminus \{v_{D-3}\}$ and covers the vertex~$v_{D-3}$. As in the first case, we add to~${\cal W}$ two watchers: $w_1 = (v_{D-2}, \{v_{D-3}, v_{D-2}, v_{D-1}\})$ and $w_2 = ( v_{D-1}, \{v_{D-2}, v_D\})$, and obtain a watching system for~$T$. So, $w(T) \leq |{\cal W}|+2 \leq w(T')+2$. The end of this case is the same as in the first case.\\

\begin{figure}
\begin{center}
\includegraphics*[scale=0.8]{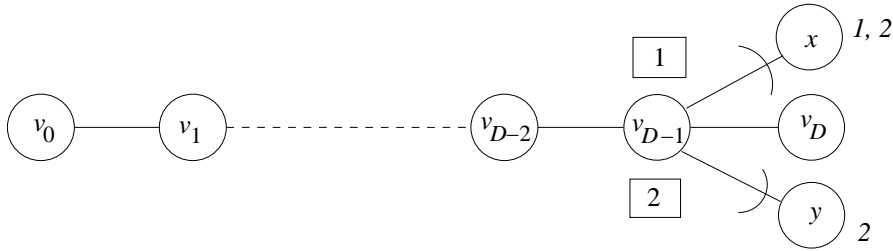}
\caption {\small Third case of Theorem~\ref{bound}: the degree of $v_{D-1}$ is at least~4.} \label{fig:thirdCase}
\end{center}
\end{figure}

\noindent $\bullet$ {\it Third case: the degree of $v_{D-1}$ is at least~4}\\
The vertex $v_{D-1}$ is adjacent to at least two vertices other than $v_{D-2}$ and~$v_D$: let $x$ and~$y$ be two neighbours of~$v_{D-1}$ distinct from $v_{D-2}$ and~$v_D$; these two vertices are leaves of $T$ (see Figure~\ref{fig:thirdCase}). We consider the tree~$T'$ obtained by removing $x$ and~$y$ from~$T$. By Lemma~\ref{lemmefeuille}, there exists a minimum watching system~${\cal W}$ of~$T'$ with a watcher~$w_1$ located at~$v_{D-1}$. For~$T$, we take the set~${\cal W}$ and add the watcher $w_2 = (v_{D-1}, \{x, y\})$; we also add the vertex~$x$ to the watching zone of~$w_1$. The set~${\cal W}$ being a watching system for~$T'$, the set ${\cal W} \cup \{w_2\}$ is a watching system for~$T$. So, $w(T) \leq w(T') + 1$.

If $n \geq 7$, the order of $T'$ is at least 5 and, using the induction hypothesis, \(w(T) \leq \frac{2}{3}(n-2)+1 < \frac{2n}{3}\).

If $n=6$, then $n-2=4$ and $w(T) \leq 3+1=4= \frac {2}{3} \times 6$.\\

\begin{figure}
\begin{center}
\includegraphics*[scale=0.8]{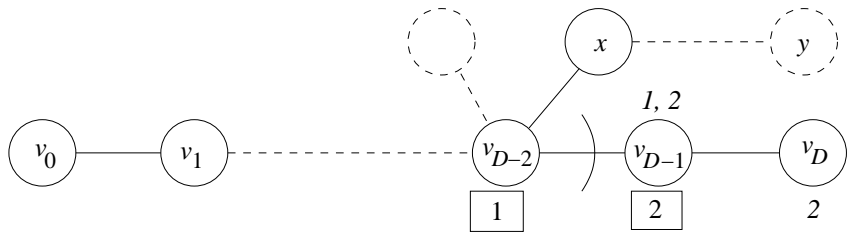}
\caption {\small Fourth case of Theorem~\ref{bound}: the degree of $v_{D-1}$ is equal to~2 and the degree of $v_{D-2}$ is at least~3.} \label{fig:fourthCase}
\end{center}
\end{figure}

\noindent $\bullet$ {\it Fourth case: the degree of $v_{D-1}$ is equal to~2 and the degree of~$v_{D-2}$ is at least~3}\\
The neighbours of $v_{D-1}$ are $v_{D-2}$ and $v_D$. The vertex $v_{D-2}$ is adjacent to $v_{D-3}$ and $v_{D-1}$ but also to at least one other vertex~$x$ (see Figure~\ref{fig:fourthCase}); if the degree of~$x$ is at least~3, using the fact that the diameter of~$T$ is equal to~$D$, we can use the first or third case to conclude, with $x$ playing the part of~$v_{D-1}$.

So, we assume that the degree of~$x$ is 1 or~2; if its degree is~2, it has a neighbour~$y$ other than~$v_{D-2}$.

We consider the tree $T'$ of order $n-2$ obtained by removing $v_{D-1}$ and $v_D$ from~$T$. By Lemma~\ref{lemmefeuille}, there exists a minimum watching system~${\cal W}$ of~$T'$ with a watcher $w_1$ located at $v_{D-2}$. For~$T$, we take the set~${\cal W}$ and add the watcher $w_2 = (v_{D-1}, \{v_{D-1}, v_D\})$; we also add the vertex~$v_{D-1}$ to the watching zone of~$w_1$. Then ${\cal W} \cup \{w_2\}$ is a watching system for~$T$.

The end of this case is exactly the same as in the previous case. \qed 

\medskip

\noindent Moreover, we can almost characterize the graphs for which this bound is tight: in~\cite{auge10}, we characterize the trees~$T$ with $n$ vertices and $w(T)=\lfloor \frac{2n}{3}\rfloor$, then we characterize the graphs~$G$ with $n$ vertices and $w(G)=\lfloor \frac{2n}{3}\rfloor$ in the cases $n=3k$, $k\geq 1$, and $n=3k+2$, $k\geq 1$; the case $n=3k+1$ is more complex, and we are only able to state a conjecture for $k\geq 6$.

\section{Watching systems in paths and cycles}
Let us call a watching system $\cw$ {\it compressed} if for every vertex $v \in V(G)$ and for every set $A$ such that $\emptyset \subsetneq A \subsetneq L_\cw (v)$, there
is $v' \in V(G)$ such that $A=L_\cw (v')$. See for example Figures~\ref{ex1watchers} and~\ref{ex2watchers}.

If a watching system $\cw$ is not compressed, then we can find $v$ and $A$ satisfying $\emptyset \subsetneq A \subsetneq L_\cw (v)$ such that $A$ is not the label of any vertex in $G$. Then if for every watcher $(x,Z)$ in $L_\cw (v) \setminus A$ we redefine this watcher by $(x,Z \setminus\{v\})$, we obtain another watching system of~$G$ where the labels of all vertices are the same as before, except for $v$ that has been assigned label~$A$. Clearly, if we do this repeatedly we get a compressed watching system of $G$ with the same size as $\cw$, and thus we can always require a watching system to be compressed.

The following lemma is easy but will prove useful:
\begin{lemma} \label{compressed}
Let $G$ be a graph and $\cw$ be a compressed watching system in $G$. Then for all $v \in V(G)$, we have:
$$2^{\card{L_\cw (v)}}-1 \leq \card{ B_G(v,2) }.$$
\end{lemma}
\noindent {\bf Proof.} Since $\cw$ is compressed, all the $2^{\card{L_\cw (v)}}-1$ nonempty labels that can be formed using the watchers in~$L_{\cw}(v)$ must be attributed to vertices in~$G$. The watchers in $L_{\cw}(v)$ having their locations in~$N_G[v]$, these labels can be attributed only inside $B_G(v,2)$. \qed

\medskip

\noindent The path $P_n$ on $n$ vertices is the graph whose vertex set is $\{1,2,\ldots,n\}$ and whose edge set is $\{\{i,i+1\} : 1 \leq i \leq n-1\}$. We prove:
\begin{theorem} \label{pathn}
For all $n \geq 1$, we have:
$$w(P_n)=\left\lceil \frac{n+1}{2}\right\rceil .$$
\end{theorem}
{\noindent \bf Proof.} First let us prove that $w(P_n) \geq \frac{n+1}{2}$. The small cases are easy to handle, and we assume that $n\geq 6$. Let $\cw$ be a minimum compressed watching system of~$P_n$ and let $i$ be such that $ 1 \leq i \leq n$. By Lemma~\ref{compressed}, since $\card{ B_{P_n}(i,2) } \leq 5$, we deduce that $|L_\cw (i)| \leq 2$. Let us show that the vertices having a label of size~$2$ can be assumed to be nonadjacent. 

First, assume that two adjacent vertices $i$ and $i+1$ have respective labels $a b$ and $c d$ where $a, b, c$ and $d$ are distinct watchers. Since $\cw$ is compressed, the four vertices around $i$ and $i+1$ must be labeled by $a$, $b$, $c$ and $d$ (and thus we must also have $i>2$ and $i+1 < n-2$). Without adding watchers, we can change~$\cw$ into a new compressed watching system where, without loss of generality, the labels from $i-2$ to $i+3$ are $a - ab - b - c - cd - d$. 

Now assume that the labels of $i$ and $i+1$ are $ab$ and~$ac$; then the vertices with labels $a$,~$b$ and~$c$ must be in $i-2$, $i-1$, $i+2$ or $i+3$. If, for instance, the labels in this order are $b - a -ab -ac - c$, then we can replace them by $b - ab - a - ac - c$. It is not difficult to see that in all cases, we can get a watching system with the same size as~$\cw$, where the vertices with labels of size~$2$ are nonadjacent. 

Let us also note that we can assume that the vertices $1$ and~$n$ do not belong to this set: for instance, if the labels of $1$,~$2$ and~$3$ are $ab - a - b$, we can replace them by $a - ab - b$, and a similar observation can be made for the vertices $n-2$, $n-1$,~$n$. 

Once $\cw$ is modified, the set of vertices with size-2 labels is an independent set in the path $2$, $3$, $\ldots, n-1$ and thus has size at most $\left\lfloor \frac{n-1}{2}\right\rfloor$, and so the set of vertices with labels of size~1, whose cardinality is the same as $\cw$, has size at least 
$$n-\left\lfloor \frac{n-1}{2}\right\rfloor,$$
which is equal to $\left\lceil \frac{n+1}{2}\right\rceil$. 

Constructions proving that $\left\lceil \frac{n+1}{2}\right\rceil$ is an upper bound are easy to find; actually it is sufficient to use identifying codes (cf.~\cite{bert04}): on the paths, watching systems are no better than identifying codes, except for $n=2$, when no identifying code exists. \hfill $\square$

\medskip 

\noindent The following result on cycles is obtained in a similar way. Let $C_n$ denote the cycle of length~$n$, with vertices $1,2, \ldots ,n,$ and edges $\{ i,i+1\}$ for $i\in \{ 1,2, \ldots, n-1\}$, and~$\{n,1\}.$
\begin{theorem}
We have $w(C_4)=3$, and for $n =3$ and all $n\geq 5$:
$$w(C_n)=\left\lceil \frac{n}{2}\right\rceil.$$
\end{theorem}
{\noindent \bf Proof.} For the lower bound, using the same argument as in the proof of Theorem~\ref{pathn}, we modify~${\cal W}$ so that the set of vertices with size-2 labels is an independent set in~$C_n$, thus having size at most $\left\lfloor \frac{n}{2}\right\rfloor$. Constructions proving the upper bound are easy to find. \qed

\medskip

\noindent If we compare to identifying codes, we can see that the cycle of length three admits no identifying code and that $i(C_4) = i(C_5) = 3$; then $i(C_n)=\frac{n}{2}$ when $n$ is even, $n\geq 6$ (see~\cite{bert04}), and $i(C_n)=\frac{n+3}{2}$ when $n$ is odd, $n\geq 7$, see~\cite{dani03}. So $i(C_n)=w(C_n)$ when $n=5$ or $n$ is even, $n\geq 4$, and $i(C_n)=w(C_n)+1$ when $n$ is odd, $n\geq 7$.

\section{Computational complexity}
Let us recall what is a {\it vertex cover} in a graph $G$. An edge $e=xy \in E(G)$ is said to be {\it covered} by a vertex $v \in V(G)$ if $v$ and~$e$ are incident, i.e., if $v=x$ or $v=y$. A {\it vertex cover} in~$G$ is a set of vertices $\cc \subseteq V(G)$ such that every edge of~$G$ is covered by a at least one element $c \in \cc$. Equivalently, $\cc$ is a vertex cover if
$$\forall e=xy \in E(G), \ \ x \in \cc \mbox{ or } y \in \cc.$$
It is well known that the problem of finding the minimum cardinality of a vertex cover in a given graph is {\it NP}-hard (see~\cite{Karp}); furthermore, it was proved in~\cite{GareyJohnsonSteiner} that this problem remains {\it NP}-hard when restricted to the class of planar graphs whose maximum degree is at most~$3$, class which we denote by~$\Pi_3$. For our proof we need to go a little further. In all graphs, a vertex of degree one is never an issue when we are looking for a vertex cover, since it is easy to prove the following lemma (see~\cite{auge09c} for instance):
\begin{lemma} \label{pi'1}
Let $G$ be a graph and $xy \in E(G)$ be an edge such that the degree of the vertex~$x$ is~1, and let $G'$ be the graph obtained by removing $x$, $y$ and all their incident edges from~$G$. Then the minimum cardinality of a vertex cover in~$G$ equals the minimum cardinality of a vertex cover in~$G'$ plus~$1$.
\end{lemma}
\noindent Let $\Pi'_3$ be the class of all planar graphs where every vertex has degree~$2$ or~$3$. In addition to the aforementioned result from~\cite{GareyJohnsonSteiner}, Lemma~\ref{pi'1} proves that the following decision problem is {\it NP}-complete:\\ 

\noindent \underline{\sc Min Vertex Cover in $\Pi'_3$}

\noindent {\sc $\bullet$ Instance:} A graph $G \in \Pi'_3$ and an integer $k$;

\noindent {\sc $\bullet$ Question:} Is there a vertex cover for $G$ with size at
most $k$ ?\\

\noindent We will use this {\it NP}-complete problem in order to study the computational complexity of the following decision problem:\\

\noindent \underline{\sc Min Watching System in $\Pi_3$}

\noindent {\sc $\bullet$ Instance:} A planar graph $G'$, with maximum degree at most~3, and an integer $k'$;

\noindent {\sc $\bullet$ Question:} Is there a watching system for $G'$ with size at
most $k'$ ?\\

\noindent We prove the following:
\begin{theorem} {\sc Min Watching System in $\Pi_3$} is {\it NP}-complete.
\end{theorem}
{\noindent \bf Proof.} Let us observe that {\sc Min Watching System in~$\Pi_3$} belongs to~{\it NP}$\,$: given a watching system, it is polynomial with respect to the size of the instance, which can be taken as the order of the graph, to compute the labels of all vertices and check that they are nonempty and distinct. Now, using a polynomial reduction from the problem {\sc Min Vertex Cover in~$\Pi'_3$}, we will show that our problem is {\it NP}-complete. 

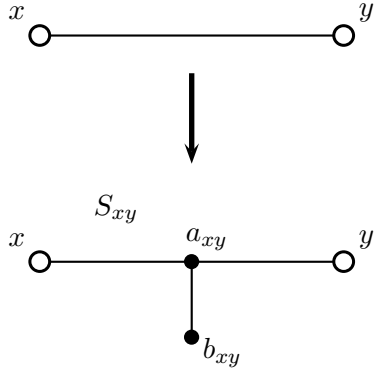
\begin{figure}[!h]
                        \centering
                        \scalebox{1}

                        \begin{pspicture}(-1,-1.5)(5,4)
                        
                        \psline(0,0)(4,0)
                        \psline(2,0)(2,-1)
			\psline(0,3)(4,3)
                        
				\psdots[dotsize=0.2](2,0)(2,-1)
                                \psdots[dotsize=0.3](0,0)(4,0)(0,3)(4,3)
                                \psdots[linecolor=white,dotsize=0.22](0,0)(4,0)(0,3)(4,3)
\rput(1,.7){$S_{xy}$}   
                        \rput(-.3,.3){$x$}    \rput(-.3,3.3){$x$} 
                        \rput(4.3,.3){$y$}       \rput(4.3,3.3){$y$}
                        \rput(2.2,.3){$a_{xy}$}
                        \rput(2.4,-1.2){$b_{xy}$}

                        \psline[linewidth=2pt]{->}(2,2.5)(2,1.3)
                        
                        \end{pspicture}
                                                                                                                                                  \caption{\small The structure $S_{xy}$ replacing every edge $xy$ of $G$ in the transformation.}  \label{structure}

                           \end{figure}

Consider a graph $G$ and an integer $k$, an instance of the {\sc Min Vertex Cover in~$\Pi'_3$} problem. Denote respectively by $n$ and~$m$ the number of vertices and edges of~$G$. We construct a graph~$G'$ by replacing every edge~$xy$ of~$G$ by the structure $S_{xy}$ depicted on Figure~\ref{structure}, consisting of 4~vertices (including~$x$ and~$y$) and 3~edges. Thus $G'$ has $n+2m$ vertices and $3m$ edges and clearly the construction of~$G'$ from~$G$ can be done in polynomial time. Moreover, if $G \in \Pi'_3$, we clearly have $G' \in \Pi_3$. We set $k'=k+m$. The reduction will be complete if we prove that for all $k \geq 0$:\\

{\it $G$ admits a vertex cover of size at most $k$ if and only if $G'$ admits a watching system of size at most~$k'$.}\\

\noindent Consider an edge $xy$ of $G$ and the structure $S_{xy}$ replacing~$xy$ in~$G'$, and let $V'_{xy}=\{a_{xy}, b_{xy}\}$.

Assume first that $\cc$ is a vertex cover of~$G$. We define a watching system~$\cw$ in~$G'$ as follows:

\begin{itemize}
\item for every vertex $x$ of $V(G)$ such that $x \in \cc$, we add the watcher $(x,N_{G'}[x])$ to~$\cw$;
\item for every edge $e=xy$ of $G$, we add the watcher $(a_{xy},N_{G'}[a_{xy}])$ to~$\cw$.
\end{itemize}

\noindent It is easy to see that $\cw$ is a watching system in~$G'$. Consider a vertex~$x$ in~$G$; since it has degree at least~2 in~$G$, it is adjacent to at least two vertices $y_1$ and $y_2$ in~$G$. So the corresponding vertex~$x$ in~$G'$ is covered by, at least, the two watchers located at~$a_{xy_1}$ and~$a_{xy_2}$, belonging respectively to the structures $S_{xy_1}$ and $S_{xy_2}$, and thus $x$~is identified by~$\cw$. Also note that for every edge $e=xy$ of~$G$, since either $x$ or~$y$ belong to the vertex cover~$\cc$, there is a watcher in~$\cw$ that separates~$a_{xy}$ from~$b_{xy}$. Thus $G'$~admits a watching system with size $|\cc|+m \leq k'$.

Conversely, assume that $\cw$ is a watching system of~$G'$ of size at most~$k'$. Consider an edge $xy \in E(G)$ and the watchers located in the structure $S_{xy}$ of~$G'$. Then:

\begin{itemize}
\item if no watcher is located at $x$ nor $y$, there must be at least two watchers located in~$V'_{xy}$;
\item if at least one watcher is located at $x$ or $y$, we still need at least one watcher in~$V'_{xy}$.
\end{itemize} 

\noindent So if we denote by $\cc$ the set of vertices $x \in V(G)$ such that $\cw$ contains a watcher located at~$x$, and by $p$ the number of edges $xy$ of $G$ with $x \not\in \cc$ and $y \not\in \cc$, we have
$$\card{\cc} \leq \card{\cw} - 2p - (m-p) \leq k' -m -p \leq k -p.$$
Therefore if we add to $\cc$ one vertex for every uncovered edge of $G$, we get a vertex cover of~$G$ of size at most $k$. \qed

\section{Distance-identification of sets of vertices}
\subsection{Definitions}
Let us now turn to the problem of identifying several vertices within a certain distance, using a watching system. For $r\geq 1$ and $ \ell \geq 1$, we define the notion of $(r,\leq \ell)$-watching systems which extends the notion of $(r, \leq \ell)$-identifying codes.

Define a $r$-watcher $w$ in a graph $G$ as in the case of a watcher $w=(v,Z)$ except for the watching zone~$Z$ that can now be any subset of the ball $B_G(v,r)$ centred at the location $v$ of~$w$ and with radius~$r$; thus a $1$-watcher is simply a watcher. We extend in an obvious way the notions of covering, label, separation, identification, $\ldots$ to $r$-watchers.

Let now $\cw$ be a set of $r$-watchers in $G$. If $A \subset V(G)$, we define the $\cw$-label of~$A$ as
$$L_\cw(A)=\bigcup_{v \in A} L_\cw(v),$$
and we say that $\cw$ is a $(r, \leq \ell)$-{\it watching system} if all the labels of the subsets~$A$ of~$V(G)$ with $1 \leq \card{A} \leq \ell$ are nonempty and distinct.

Note that a $(r,\leq \ell)$-watching system is a $(r',\leq \ell')$-watching system if $\ell' \leq \ell$ and $r' \geq r$.

Let ${\cal S}=\{S_1, S_2,\ldots,S_k\}$ be a finite family of distinct nonempty subsets of a set~$X$ and $\ell \geq 1$. We say that $\cal S$~is a $\ell$-{\it superimposed family} on~$X$ if, whenever we consider two distinct sets $I, J$ included in $\{1,\ldots,k\}$ with $1 \leq \card{I}\leq \ell$ and $1 \leq \card{J}\leq \ell$, we have:
$$\bigcup_{i \in I} S_i \neq \bigcup_{j \in J} S_j.$$
This is the notion of $\ell$-superimposed code in a set-system version. These codes were introduced in~\cite{kaut64}. They are related to $(r, \leq \ell)$-identifying codes, as was observed in~\cite{karp98a} and~\cite{frie06}. They are also related to watching systems since, with our definition, if $\cw$ is a $(r,\leq \ell)$-watching system in a graph~$G$, then the family of all $\cw$-labels of the vertices of~$G$ is a $\ell$-superimposed family on~$\cw$. Note that the family of singletons of~$X$ is always a $\ell$-superimposed family of~$X$ for all $\ell \geq 1$, and so every graph~$G$ admits a $(r, \leq \ell)$-watching system for all $r \geq 1$ and $\ell \geq 1$, consisting of the watchers $(v,\{v\})$ for all $v \in V(G)$. 

Observe that if $\ell \geq 2$ and $i\neq j$, then $S_i \subseteq S_j$ is impossible in a $\ell$-superimposed family. {F}rom this follows that if $\card{L_\cw(x)}=1$ for a vertex~$x$ in the graph with watching system~$\cw$, then if $ \ell \geq 2$ the watcher covering~$x$ must cover only~$x$: we will call such a watcher a {\it hermit}. Without loss of generality, we can suppose that this watcher is $(x,\{x\})$, since its location does not matter.

\subsection{The case of $(1,\leq 2)$-watching systems in paths and cycles}
\label{subsec12pc}
Let us start with the following lemma.
\begin{lemma} \label{superimposed}
For $1 \leq k \leq 4$, the only $2$-superimposed family on a set with $k$ elements
with at least $k$ subsets is the family of $k$ singletons.
\end{lemma}
{\noindent \bf Proof.} The result is obvious if $1 \leq k \leq 3$, so we just check the case $k=4$. Let $S_1$, $S_2$, $S_3$, $S_4$ be a $2$-superimposed family on $\{1,2,3,4\}$. If there is a singleton in the family, say $S_1=\{1\}$, then we have $S_i \subset \{2,3,4\}$ for $i>1$ and we use the case $k=3$ to conclude.

If an element, say $1$, is in at least three different sets, say $S_1$, $S_2$ and~$S_3$, then by intersecting these sets with $\{2,3,4\}$ we get a $2$-superimposed family of size~3 on $\{2,3,4\}$, so, using the case $k=3$, we must have (up to permutations) $S_1=\{1,2\}$, $S_2=\{1,3\}$ and $S_3=\{1,4\}$. Then $S_4$ cannot contain~$1$, and the remaining possibilities for~$S_4$ all lead to contradictions.

If all the elements are in at most two sets and there are no singletons, then by a simple counting argument we see that all the sets must be pairs, and so the family must be (up to permutations) $\left\{\{1,2\} , \{1,3\}, \{3,4\}, \{2,4\}\right\}$, which is not $2$-superimposed. \hfill $\square$

\medskip

\noindent In other words, with $k$ watchers, $1 \leq k \leq 4$, we can produce $k$~valid labels, which will be singletons, and not more.

{F}rom now on until the end of Section~\ref{subsec12pc}, the vertices of the paths or cycles are denoted by $x_1, x_2, \ldots$ and the watchers by $1, 2, \ldots$ or $w_1, w_2, \ldots$, depending on the context.
\begin{theorem} \label{frac56}
For all $n \geq 1$, the minimum size of a $(1, \leq 2)$-watching system in the path~$P_n$~is
$$\min(n,\Big \lceil \frac{5(n+1)}{6} \Big \rceil ), \mbox{ which is equal to } \: \: \left \{ 
\begin{array}{ll}
n & \mbox{ if } n \leq 10,\\
\Big \lceil \frac{5(n+1)}{6} \Big \rceil & \mbox{ if } n \geq 5.
\end{array}
\right. $$
\end{theorem}
{\noindent \bf Proof.}
Let $\lambda _n=\big \lceil \frac{5(n+1)}{6} \big \rceil$ and $\Lambda _n=\min(n,\lambda _n)$: we have to prove that $\Lambda _n$ is the minimum size of a $(1, \leq 2)$-watching system in~$P_n$. The proof for the lower bound works by induction on~$n$.

If $1\leq n \leq 4$, the result comes directly from Lemma~\ref{superimposed}: we need $n$ watchers since the labels of the vertices form a $2$-superimposed family of size~$n$. So from now on, we assume that $n\geq 5$, that the lower bound holds for all $n'<n$, and that ${\cal W}$ is a $(1,\leq 2)$-watching system in~$P_n$. We call a {\it hole} any vertex at which no watcher is located.

If  there is no hole, then $|{\cal W}| \geq n$, and so $|{\cal W}| \geq \Lambda _n$ for all $n\geq 5$; therefore, we can assume that at least one hole exists in~$P_n$.

Case~1) Suppose first that holes exist only on $x_1$ or $x_n$.\\
Without loss of generality, we can assume that $x_1$ is a hole. If $x_1$ is covered by only one watcher~$w$, this watcher is located at~$x_2$ and cannot cover any other vertex~$x_i$, $i\in\{2,3\}$, since this would imply that the sets $\{x_1,x_i\}$ and $\{x_i\}$ cannot be separated by~${\cal W}$. So $w=(x_2,\{x_1\})$; then
$${\cal W}\,'=\big ({\cal W} \setminus \{w\}\big ) \cup \{(x_1,\{x_1\})\}$$
is still a $(1,\leq 2)$-watching system in~$P_n$, and 
$${\cal W}\, ''={\cal W}\,' \setminus \{(x_1,\{x_1\})\}$$
is a $(1,\leq 2)$-watching system in~$P_{n-1}=x_2,x_3, \ldots ,x_n$. By the induction hypothesis, we have $|{\cal W}\,''|\geq \Lambda _{n-1}$, from which we immediately derive that $|{\cal W}| = |{\cal W}\,''| +1 \geq \Lambda _{n}$.

If now $x_1$ is covered by at least two watchers $w_i$, $i=1,2, \ldots$, located at~$x_2$, then we consider the path $x_3, x_4, \ldots, x_n$ and the set ${\cal W}_1={\cal W} \setminus \{w_i : i=1,2, \ldots\}$. We then have three subcases: 

$\bullet$ if $x_n$ is not a hole, then $x_1$ is the only hole in~$P_n$, and $|{\cal W}_1|\geq n-2$, implying $|{\cal W}| \geq n \geq \Lambda _n$; 

$\bullet$ if $x_n$ is a hole covered by only one watcher, we use the same argument as used in this case for~$x_1$, and we are done; 

$\bullet$ if $x_n$ is a hole covered by at least two watchers $w'_j$, $j=1,2, \ldots$, located at~$x_{n-1}$, then we consider the path $P_{n-4}=x_3, x_4, \ldots, x_{n-2}$ and the set ${\cal W}_2={\cal W}_1 \setminus \{w'_j : j=1,2, \ldots\}$; by hypothesis, $P_{n-4}$ has no holes, so $|{\cal W}_2| \geq n-4$ and again $|{\cal W}| \geq n \geq \Lambda _n$. In all three cases, we are done, which ends Case~1.

Case~2) There is a hole $x_{\ell}$, $\ell \in \{2,3, \ldots, n-1\}$.\\
Without loss of generality, we can assume that $\ell \leq \lfloor \frac{n+1}{2} \rfloor$, so that
 there are more vertices to the right of~$x_{\ell}$ than to its left: $\ell-1\leq n-\ell$. We distinguish between three subcases:

$\bullet$ $\ell -1 \leq 4$ and there are at least $\ell$ watchers located at $x_1, x_2, \ldots, x_{\ell -1}$; since $r=1$, there is no influence by these $\ell$~watchers on~$x_{\ell +1}$, and we can apply the induction hypothesis to $x_{\ell +1}, \ldots , x_n$, and obtain:
$$|{\cal W}| \geq \ell +\Lambda_{n-\ell} = \ell + \min \big (n-\ell, \Big \lceil \frac{5(n-\ell +1)}{6} \Big \rceil \big )=\min \big (n, \Big \lceil \frac{5(n+1)+\ell}{6} \Big \rceil \big ) \geq \Lambda _n.$$

$\bullet$ $\ell -1 \leq 4$ and there are exactly $\ell -1$ hermits located at $x_1, x_2, \ldots, x_{\ell -1}$ (by Lemma~\ref{superimposed}, and without loss of generality as far as the hermits' locations are concerned, this is the only possibility left when $\ell -1 \leq 4$); since these hermits cannot interfere with~$x_{\ell}$, we can apply the induction hypothesis to $x_{\ell}, \ldots , x_n$, and obtain this time:
$$|{\cal W}|\geq \ell -1+\Lambda_{n-\ell +1} = \ell -1+ \min \! \big (n-\ell +1, \! \Big \lceil \frac{5(n-\ell +2)}{6} \Big \rceil \big )\! =\min \! \big (n, \! \Big \lceil \frac{5(n+1)+\ell -1}{6} \Big \rceil \big )\! \geq \Lambda _n.$$

$\bullet$ $\ell -1 \geq 5$ (and so, $n-\ell \geq 5$); again, $x_{\ell-1}$ cannot interfere with $x_{\ell+1}$, and we can apply the induction hypothesis to $x_1, x_2, \ldots , x_{\ell -1}$ and to $x_{\ell +1}, \ldots, x_n$, and obtain:
$$|{\cal W}|\geq \Big \lceil \frac{5\ell}{6} \Big \rceil + \Big \lceil \frac{5(n-\ell+1)}{6} \Big \rceil \geq \Big \lceil \frac{5(n+1)}{6} \Big \rceil \geq \Lambda _n,$$
which ends the proof for the lower bound.

\begin{figure}[!h]
                        \centering
                        \scalebox{1}

                        \begin{pspicture}(+2,-2)(10,2)
                        
                        \psline(1,0)(11,0)                      
\psdots[dotsize=0.2](1,0)(2,0)(3,0)(4,0)(5,0)(6,0)(7,0)(8,0)(9,0)(10,0)(11,0)
                         \rput(1,.6){\psframebox[fillstyle=solid,fillcolor=white]{$1$}}
                        \rput(2,.6){\psframebox[fillstyle=solid,fillcolor=white]{$2$}}
                        \rput(3,.6){\psframebox[fillstyle=solid,fillcolor=white]{$3$}}
                        \rput(4,.6){\psframebox[fillstyle=solid,fillcolor=white]{$4$}}
                        \rput(5,.6){\psframebox[fillstyle=solid,fillcolor=white]{$5$}}
                        \rput(7,.6){\psframebox[fillstyle=solid,fillcolor=white]{$6$}}
                        \rput(8,.6){\psframebox[fillstyle=solid,fillcolor=white]{$7$}}
                        \rput(9,.6){\psframebox[fillstyle=solid,fillcolor=white]{$8$}}
                        \rput(10,.6){\psframebox[fillstyle=solid,fillcolor=white]{$9$}}
                        \rput(11,.6){\psframebox[fillstyle=solid,fillcolor=white]{$10$}}
                         \rput(1,-.4){\small \it 1, 2}
                        \rput(2,-.8){\small \it 1, 3}
                        \rput(3,-.4){\small \it 2, 4}
                        \rput(4,-.8){\small \it 3, 5}
                        \rput(5,-.4){\small \it 4, 5}
                        \rput(6,-.8){\small \it 5, 6}
                        \rput(7,-.4){\small \it 6, 7}
                        \rput(8,-.8){\small \it 6, 8}
                        \rput(9,-.4){\small \it 7, 9}
                        \rput(10,-.8){\small \it 8, 10}
                        \rput(11,-.4){\small \it 9, 10}
                        
                        \end{pspicture}
\caption{\small An optimal ($1, \leq 2$)-watching system in the path $P_{11}$.}   
\label{path}

            \end{figure}
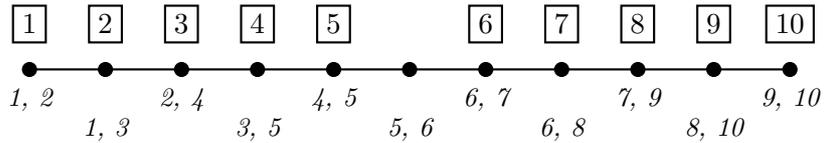

We give a construction that matches the lower bound. For $n\in \{1,2, \ldots, 10\}$, we need at least $n$ watchers, and we can do it with $n$ hermits. For $n=11$, see Figure~\ref{path}. 

When $n=6k-1$, $k\geq 3$, for which at least $5k$ watchers are necessary, use Figure~\ref{path}, add the pattern of the last six vertices to the right of the right-most vertex, $x_{11}$, change, for these new vertices, $6,7,8,9,10$ into $11,12,13,14,15$, and so on.

When $n=6k+i$, $k\geq 2$, $i\in \{0,1,2,3,4\}$, for which at least $5k+i+1$ watchers are necessary, use the construction for $6k-1$, where the watchers $5k-4, \ldots , 5k$ are located at the last five vertices, and where the label of the last vertex, $x_{6k-1}$, is $\{5k-1,5k\}$. Now add $n-6k+1=i+1$ vertices to the right of $x_{6k-1}$, and locate $i+1$ watchers $5k+1, \ldots , 5k+i+1$ at the vertices $x_{6k}, \ldots, x_{6k+i}$. Change the label of $x_{6k-1}$ to $\{5k-1,5k+1\}$ and assign the labels $\{5k+j,5k+j+2\}$ to $x_{6k+j}$, for $j\in \{0, \ldots, i-1\}$, and the label $\{5k+i,5k+i+1\}$ to $x_{6k+i}=x_n$.

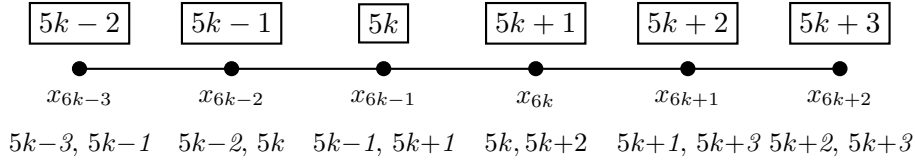
\begin{figure}[!h]
                        \centering
                        \scalebox{1}

                        \begin{pspicture}(+2,-2)(10,2)
                        
                        \psline(1,0)(11,0)                      
\psdots[dotsize=0.2](1,0)(3,0)(5,0)(7,0)(9,0)(11,0)
                         \rput(1,.6){\psframebox[fillstyle=solid,fillcolor=white]{$5k-2$}}
                        \rput(3,.6){\psframebox[fillstyle=solid,fillcolor=white]{$5k-1$}}
                        \rput(5,.6){\psframebox[fillstyle=solid,fillcolor=white]{$5k$}}
                        \rput(7,.6){\psframebox[fillstyle=solid,fillcolor=white]{$5k+1$}}
                        \rput(9,.6){\psframebox[fillstyle=solid,fillcolor=white]{$5k+2$}}
                        \rput(11,.6){\psframebox[fillstyle=solid,fillcolor=white]{$5k+3$}}
                         \rput(1,-.4){\small $x_{6k-3}$}
                         \rput(3,-.4){\small $x_{6k-2}$}
                         \rput(5,-.4){\small $x_{6k-1}$}
                         \rput(7,-.4){\small $x_{6k}$}
                         \rput(9,-.4){\small $x_{6k+1}$}
                         \rput(11,-.4){\small $x_{6k+2}$}
                         \rput(1,-1){\small $5k-${\it 3}, $5k-${\it 1}}
                        \rput(3,-1){\small $5k-${\it 2}, $5k$}
                        \rput(5,-1){\small $5k-${\it 1}, $5k+${\it 1}}
                        \rput(7,-1){\small $5k, 5k+${\it }2}
                        \rput(9,-1){\small $5k+${\it 1}, $5k+${\it 3}}
                        \rput(11,-1){\small $5k+${\it 2}, $5k+${\it 3}}
                        
                        \end{pspicture}
\caption{\small The right end  of the path $P_n$ for $n=6k+2$.}   
\label{pathbisAL}

            \end{figure}

We let the reader check that this is indeed a $(1,\leq 2)$-watching system; see Figure~\ref{pathbisAL} for an example. \qed

\medskip

\noindent Observe also that no $(1, \leq 2)$-identifying code (and more generally, no $(1,\leq \ell)$-identifying code) exists in the path~$P_n$, because, since $N_{P_n}[x_1] \subseteq N_{P_n}[x_2]$, the sets of vertices $\{x_2\}$ and $\{x_1,x_2\}$ cannot be separated.

In the case of cycles, we have the following result.
\begin{theorem}
For all $n \geq 3$, the minimum size of a $(1, \leq 2)$-watching system in the cycle~$C_n$ is $\lceil \frac{5}{6} n \rceil$, except for $n=6$, for which it is~6.
\end{theorem}
{\noindent \bf Proof.} 
The small cases, up to $n=5$, are easy to handle. The proof of the case $n=6$ is cumbersome and is not given here. Now consider a $(1, \leq 2)$-watching system~${\cal W}$ in~$C_n$, $n\geq 7$. Remembering that a hole is a vertex at which no watcher is located, we can see that if there is no hole, we are done; so we consider a hole~$x_{\ell}$ in~$C_n$, and the path on $n-1$ vertices obtained from~$C_n$ by removing~$x_{\ell}$. Obviously, ${\cal W}$ is a $(1, \leq 2)$-watching system for this path, and so $|{\cal W}| \geq \Lambda _{n-1}=\lceil \frac{5((n-1)+1)}{6} \rceil$, and the lower bound is proved. Constructions meeting the lower bound are easy: take a path with $n-1$ vertices together with the construction of an optimal $(1, \leq 2)$-watching system described in the proof of Theorem~\ref{frac56}. Add a vertex~$x_n$ which is linked to~$x_1$ and~$x_{n-1}$, and assign to~$x_n$ the label $\{w_1,w_2\}$, where $w_1$ is located at~$x_1$ and $w_2$ is located at~$x_{n-1}$ (in our construction, there are always watchers located at each end of the path). You obtain a $(1, \leq 2)$-watching system for~$C_n$, of size $\Lambda _{n-1}=\lceil \frac{5n}{6} \rceil$, for $n\geq 7$. The reason why this construction does not work for $n=6$ is that there would be three labels, $\{1,3\}$, $\{3,5\}$ and $\{1,5\}$, whose pairwise unions are equal to $\{1,3,5\}$. \qed

\medskip

\noindent If we compare to $(1, \leq 2)$-identifying codes in $C_n$, we can see that, because for all~$i$ the sets $B_{C_n}(x_i,1)$ and $B_{C_n}(x_i,1) \cup B_{C_n}(x_{i+1},1)$ differ by only one vertex, $x_{i+2}$, this vertex, hence by symmetry all vertices, must belong to the code. Starting from $n=7$, the only $(1, \leq 2)$-identifying code in the cycle~$C_n$ is~$V(C_n)$.
\subsection{The case of $(1,\leq \ell)$-watching systems in paths and cycles for $\ell \geq 3$}
Like every graph, the path $P_n$ and the cycle $C_n$ admit, for all $\ell \geq 3$, a $(1, \leq \ell)$-watching system, which is the trivial watching system consisting of all the hermits. In the case of~$P_n$ and~$C_n$, this is the best we can do:
\begin{theorem}
For all $n \geq 1$ (respectively, $n\geq 3$) and $\ell \geq 3$, the minimum size of a $(1, \leq \ell)$-watching system in the path~$P_n$ (respectively, the cycle~$C_n$) is~$n$.
\end{theorem}
\noindent {\bf Proof.} Consider a $(1, \leq \ell)$-watching system $\cw$ for $P_n$ or~$C_n$, where $\ell \geq 3$. Let $\cal H \subseteq \cw$ be the set of hermits in~$\cw$, and let $V_{\cal H} \subset V(G)$ be the set of vertices covered by these hermits (as aforementioned, $V_{\cal H}$ can be taken, without loss of generality, as the set of the locations of the hermits). Now assume that there is a vertex~$x$ in $V(G) \setminus V_{\cal H}$; then we have $|L_\cw(x)|>1$.

Suppose that $L_\cw(x)=\{w_1,w_2\}$: then $w_1$ and $w_2$ are not hermits and so there exist a vertex $v_1 \neq x$ covered by~$w_1$ and a vertex $v_2 \neq x$ covered by~$w_2$ (though we may have $v_1=v_2$). But in this case we have $L_\cw(\{v_1,v_2\})=L_\cw(\{v_1,v_2,x\})$ (or $L_\cw(\{v_1\})=L_\cw(\{v_1,x\})$ if $v_1=v_2$), and so $\cw$ cannot be a $(1, \leq \ell)$-watching system if $\ell \geq 3$.

Therefore, all vertices in $V(G) \setminus V_{\cal H}$ are covered by at least three watchers from $\cw \setminus \cal H$; since a watcher can only cover at most three vertices, we clearly have $\card{\cw \setminus \cal H} \geq \card{V(G) \setminus V_{\cal H}}$. Since we also have $\card{\cal H}=\card{V_{\cal H}}$, the result follows. \qed

\medskip

\noindent Finally, observe that for $\ell \geq 3$, no $(1, \leq \ell)$-identifying code exists in the cycle~$C_n$, because the sets of vertices $\{1,3\}$ and $\{1,2,3\}$ (or more generally, $\{x,x+2\}$ and $\{x,x+1,x+2\}$) cannot be separated.

\end{document}